# A Conversation with Murray Rosenblatt

**David R. Brillinger and Richard A. Davis**

**In Memory of Ady Rosenblatt 1926-2009.**

*Abstract.* On an exquisite March day in 2006, David Brillinger and Richard Davis sat down with Murray and Ady Rosenblatt at their home in La Jolla, California for an enjoyable day of reminiscences and conversation. Our mentor, Murray Rosenblatt, was born on September 7, 1926 in New York City and attended City College of New York before entering graduate school at Cornell University in 1946. After completing his Ph.D. in 1949 under the direction of the renowned probabilist Mark Kac, the Rosenblatts' moved to Chicago where Murray became an instructor/assistant professor in the Committee of Statistics at the University of Chicago. Murray's academic career then took him to the University of Indiana and Brown University before his joining the University of California at San Diego in 1964. Along the way, Murray established himself as one of the most celebrated and leading figures in probability and statistics with particular emphasis on time series and Markov processes. In addition to being a fellow of the Institute of Mathematical Statistics and American Association for the Advancement of Science, he was a Guggenheim fellow (1965–1966, 1971–1972) and was elected to the National Academy of Sciences in 1984. Among his many contributions, Murray conducted seminal work on density estimation, central limit theorems under strong mixing, spectral domain methods and long memory processes. Murray and Ady Rosenblatt were married in 1949 and have two children, Karin and Daniel.

## WHEN MURRAY MET ADY

**Richard:** Ady, how did you and Murray meet?

**Ady:** We met in the Fordham Road library. He came over and we talked and went for a walk in the pouring rain. We walked over to his girlfriend's house. She wasn't home, so we left a message with her sister or whatever. The next day his girlfriend said to me "Oh, that was interesting you dropped over the same day that Murray did." And I said "Oh yeah, we were together." Things went down after that. I guess we were married when we were around 23 in 1949. I was older, I was 23 and he was still 22. We were in Ithaca. I waited until he finished graduate school.

*David Brillinger is Professor of Statistics, Department of Statistics, University of California, Berkeley, CA 94720-3860, USA e-mail: brill@stat.berkeley.edu.
Richard Davis is Howard Levene Professor of Statistics, Department of Statistics, 1255 Amsterdam Avenue, Columbia University, New York, NY 10027, USA e-mail: rdavis@stat.columbia.edu.*

**Richard:** I guess he was a little slow! Were you attending Cornell at the same time?

**Ady:** No, I had a job teaching swimming in NYC. We saw each other more or less during the war and then he went up to Cornell and I would go up and visit him every now and then.

**Richard:** So you met in high school.

**Ady:** No, we met after high school.

**Richard:** But you graduated when you were 16?

**Ady to Murray:** Weren't we 16 when we graduated?

**Murray:** Probably.

**Ady:** Maybe I was 17, I don't know because I was born in April. I don't remember when I graduated. I remember when I got married!

**David:** Do you remember the best man and maid of honor at your wedding?

**Ady:** I think it was Bert and Shirley Yood.

**Murray:** Bert Yood was a specialist on Banach algebras. I remember taking a course from him on Banach algebras as a graduate student.

**David:** Were your parents and his parents immigrants?

**Ady:** No, his parents were immigrants. His father was from Russia (now Ukraine) and his mother was from Poland. They met in the States. After we walked home from Murray's girlfriend's house in the rain, he caught a cold. His mother was not happy with me when she met me because I gave him that.

**Richard:** So you didn't make a good first impression!

**David:** How about your parents?

**Ady:** My father's parents came from Poland. He and his younger sister were born in America. The rest of his siblings were born in Poland. And my mother's family came from Hungary and Austria. My grandfather was Hungarian. They were all Jewish.

**Richard:** Murray had one sibling?

**Ady:** Yes, Murray had one brother David, who was always a major influence on him.

**Murray:** Oh, yes. Although I didn't always follow his advice, I found it useful to listen to.

### ATTENDING CCNY AND CORNELL

**Richard:** Murray, back at CCNY, one of your professors, Emil Post, seemed to have made quite an impact on you.

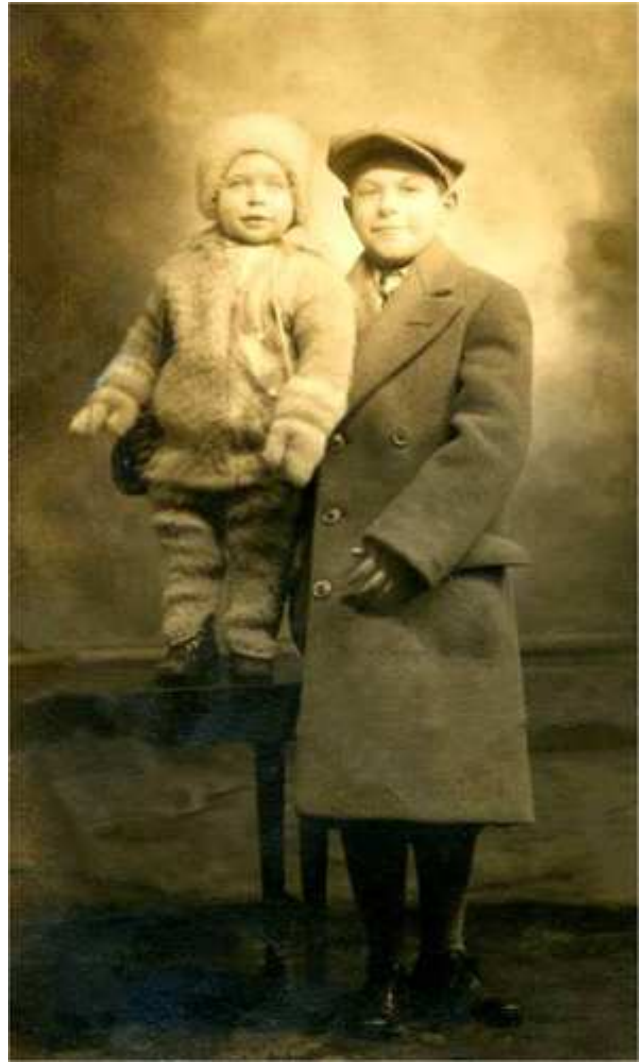

FIG. 2. *Murray with his older brother David around 1928.*

**Murray:** Actually, Post was a remarkable character. I think he is, at least on the American scene, maybe internationally, one of the great figures in mathematical logic. Because he was manic-depressive he used to get into these manic states occasionally and had to be institutionalized. Also he was one-armed. He actually did some real analysis too. I took a class in real analysis with him. He was following some book with an incredible number of errors, which he corrected. And he was sort of a perfectionist. There's an amusing story. Martin Davis, a fellow student, who's a well-known figure in mathematical logic today who has done some remarkable work, was also in the class. Post used to assign problems and have people come up to the board. I guess at one point he asked me and I guess I was starting in a particular direction and he was about to cut me off when Davis said why don't you let him go on, which you might say saved me at that point. But I took a reading course with him later on, this person who was so formal in the class, turned out to be a very pleasant human. He used a book, actually a very lovely book by a French mathematician by the name of C. de la Vallée Poussin. The book [18] was his *Intégrales de Lebesgue, Fonctions d'Ensemble, Classes de Baire* published in 1916. I may have a copy of it. It was actually a very elegant book and it was a pleasant reading course to take with him, and have occasionally some interchanges with him. I have rather pleasant memories of that.

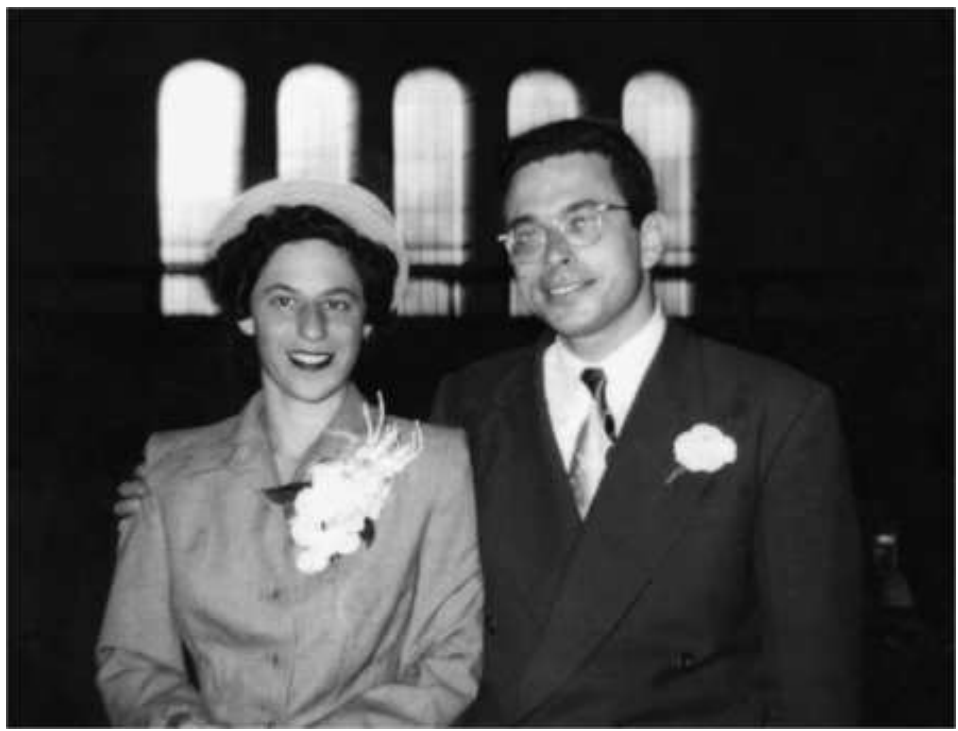

Fig. 1. *Wedding photo (1949).*

**David:** Was he an American?

**Murray:** Oh yes, he was American. Probably at that time he was the most distinguished member of the CCNY mathematics faculty. Unfortunately, at one point I came into class and he was enthusiastic. That was a sign that he was entering a manic phase. They took him off to treatment I assume.

**Richard:** What other courses did you take at CCNY? I was just wondering how you became interested in probability and statistics.

**Murray:** I remember taking courses at City College in mathematical physics and thermodynamics. I probably didn't have any interest in probability and statistics then. I went to Cornell as a graduate student and William Feller and Mark Kac were on the faculty. I took most of the courses in probability theory with Feller. I wrote my thesis with Mark Kac as my advisor. Feller was a remarkable mathematician but had very strong but humorous opinions and great enthusiasm. A fellow student, Samuel Goldberg, and I used to take notes on Feller's lectures. Feller thanked him in the introduction in his well-known book on probability theory, the lovely book on introductory probability theory. At the beginning of a discussion of the 3 series theorem, Feller said isn't it obvious? I guess some of us had enough self-preservation or ego to say no, we don't see it's obvious. It took two to three lectures to go through the full development.

**Richard:** You went to Cornell with the idea of studying mathematics with no particular specialty in mind?

**Murray:** Yes, it was mathematics. I guess there were two opportunities at that time, either Brown or Cornell. For whatever reasons, I chose Cornell.

**Ady:** Didn't they give you a better salary?

**Murray:** I'm sure they did. There were younger people with a good deal of interest in probability theory too. Gilbert Hunt was there and so was Kai

Lai Chung. And in one year there were quite a number of visitors, Doob, Donsker, Darling and various of Doob's students, I guess. There must have been Laurie Snell and John Kinney. So there was a good deal of activity in that area. I guess I probably took a course in mathematical statistics and I suppose it was given by Feller. It was good as a student to be working on a thesis under Kac—you knew with Kac, you could come in and talk with him if you wanted to and you would get good advice. He didn't have strong opinions about this being the direction to go into while Feller did have such propensities. Mario Juncosa was a graduate student with me there. We still keep in contact with him; he has been at Rand for many years and Juncosa was a student of Feller's. He completed his thesis there.

**David:** Ady, did you get to know Marc Kac? What did you think of him?

**Ady:** I thought he was a lovely person. He was very, very kind and very nice and helpful. His wife was Kitty. We saw them through the years actually.

**Murray:** He moved to California, to USC.

**Ady:** He was at Rockefeller before. We used to see him there at USC. Shortly before he died we used to go up to the theater with them. He was really nice.

**Richard:** Were graduate students supported as teaching assistants and such?

**Murray:** The first year I had an Erastus Brooks fellowship at Cornell. The second year I taught classes. The Office of Naval Research came through with support in the last year. So I was supported, initially on a fellowship, then what amounted to assistantships.

**Ady:** Didn't ONR support you with grants all the way through your career?

**Murray:** A good deal of the time, but I also had partial support from NSF. Certainly, the Office of Naval Research supported me.

**Richard:** I suppose I was also supported at some point on Murray's ONR grants during my graduate studies.

**Murray:** I think a good many of the students I had were supported by NSF too but mainly the Office of Naval Research.

**Richard:** At Cornell did they support you directly or was it funneled through a faculty member?

**Murray:** It must have funneled through a faculty member, maybe Mark Kac.

**David:** I picture you as interested in applications in the physical sciences. Did that start when you went to Brown or were you doing that at Cornell also?

**Murray:** Just in terms of the background at Cornell, Kac always had interest in those parts of physics related to statistical mechanics so there's definitely in the background an interest in applications. A good deal of probability theory initially was motivated by applications of sorts, maybe initially to gambling systems but other areas too. I suppose some of it may relate to undergraduate courses at CCNY in mathematical physics and thermodynamics. Well, you know, my thesis... what was the title of my thesis?

**Richard:** I have it here, Murray, in case you can't remember.

**Murray:** Maybe something on Wiener functionals.

**Richard:** "On distributions of certain Wiener functionals."

**Murray:** Right, and that was an attempt to mildly generalize some results of Kac; You know, this is really related to what was later referred to as the Kac-Feynman formula. A revised version of it is published in a paper [19] called "On a class of Markov processes" which appeared in the Transactions of the American Mathematical Society. What it does is to consider the integral of some function of both time and Brownian motion. What one essentially does is look at the Laplace transform of the fundamental conditional distribution and relate that to a solution of an associated parabolic differential equation. So at least formally, that's related to the formula.

**Murray:** Actually, I wrote a master's thesis and it was never published. My thesis was on definitions of absolute continuity for functions of two variables. As I remember, my thesis committee included two members. One of the members, I'm trying to remember, in his own day, was a well-known mathematician called Wally Hurwitz.

**Ady:** Oh, right. He was good at the stock market, right?

**Murray:** Oh yes, he knew how to invest in the stock market and he left quite a bit to Cornell University, to the Mathematics Department. But he was very good. My doctoral thesis committee included three people. I remember Mark Kac of course, who was my thesis advisor and Morrison.

**David:** Which Morrison?

**Murray:** The Philip Morrison who retired eventually from MIT. He is a well-known name in physics. I'm not sure I still have it [my thesis]. It may be at school or I might have lost it. It's possible Harry Pollard was a part of that committee too, but I don't know. If they were part of that committee they may have helped me through because I'm sure some of my answers to the physically oriented questions of Morrison may not have been that adequate. Actually, I had what was called a minor in physics which consisted of a series of courses in quantum mechanics taught by Hans Bethe.

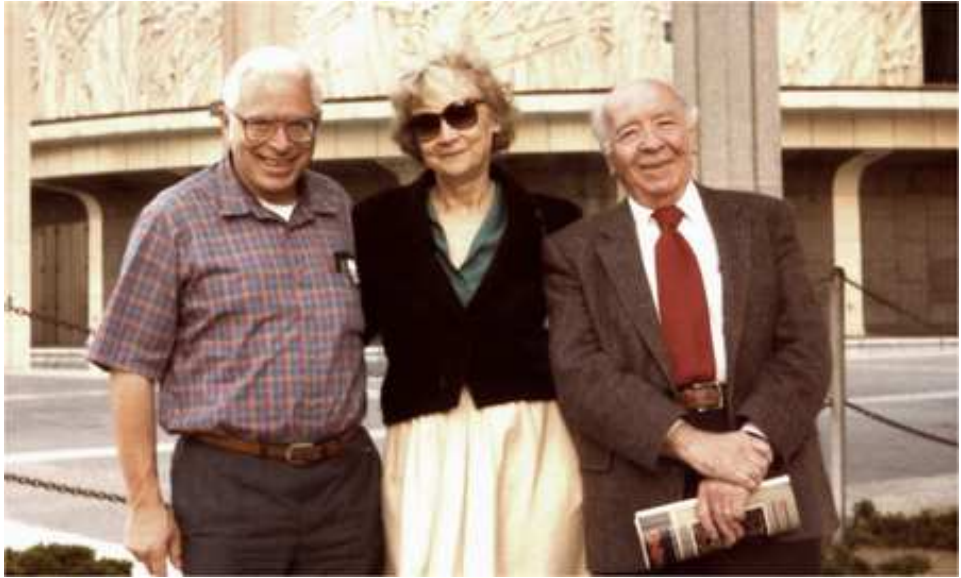

Fig. 3. *Murray with Katherine and Mark Kac (1983).*

**David:** Wow! And Feynman was there too.

**Murray:** Oh yes. Feynman was a hilarious character.

**Richard:** What was that like? I guess there was one symposium with him and Feller going at it.

**Murray:** There was a lecture. No, not a lecture, but an interchange with Feynman, Kac and Feller. Feynman had such agility, in terms of instantaneous and spontaneous response. He put both of them to shame. The one person you can compare him to, and I think he was even better, was the prime conservative, Milton Friedman. I thought Feynman was even better than Friedman. Friedman was probably the most articulate defender of the conservative perspective in economics.

**David:** I guess in World War II he was a part of that statistics group at Columbia.

**Murray:** I think Hotelling started out as the head of that group. It ended up with Allen Wallis as head. Probably due to Wallis' abilities administratively, I don't know. So from that point of view, I guess you might say my interest in physics partially comes from my graduate student days since I was exposed to some of it then.

**Richard:** You finished in three years; that seems incredibly fast to me, especially if one includes a Master's thesis on top of that.

**Murray:** Kac must have been the reason. I am eternally indebted to Kac as the person who served as a thesis advisor and may have helped occasionally with suggestions, but sort of left you alone without saying you've got to do this or that so forth and so on. He let you to go your own way.

**Richard:** This sounds familiar actually.

**Murray:** In what sense? In your case, I didn't have to give any direction anyways. What did I do? I suggested an area, you got into it and worked on it.

**Richard:** Was that your method of operation in advising students?

**Murray:** Well, it seemed to me if a student is bright enough to make his own way, why do you have to impose on him?

**David:** The range of their thesis topics is very broad.

**Murray:** Well, one student I got into a field was Richard Bradley and obviously he continued. He has become the great expert on strong mixing. Look at his marvelous three volumes on strong mixing conditions [3].

**Richard:** It always seems like you had a hidden motive in mind regarding the topics that we worked on. When I was a graduate student, Rick Bradley was working on strong mixing, Ed Mack was working on density estimation and I was working on extremes under mixing conditions and it seemed like....

**Murray:** You were also looking at some aspects of Markov processes, relative to extremes, right?

**Richard:** Yes, but the main topic was extremes of stationary processes and it seemed that you had some other application in mind. There was a connection between these components that *you* saw but was invisible to us.

**Murray:** I don't think it was anything that conscious. You know what I was doing. If a student wanted a thesis topic and I hadn't thought of one, it seemed reasonable or interesting to suggest things that I had marginal acquaintance with that sounded interesting and look if it's possible to work on.

**Richard:** Later, I could see the connection with density estimation and the results you had with Bickel on maximum deviation of density estimates.

**Murray:** The density estimation actually comes out of the spectral estimation in a direct manner. It is sort of silly because it's obvious. It's an example how contrary to the usual notion it is to do things in a simpler situation and then go on to greater complexity. I mean, what happened in the density function estimation was that I had certain results on spectral estimation. I saw the paper of Fix and Hodges [10], and in my paper on density estimation that paper is referred to. The notion was good. They proposed some density estimate, and the notion was to look at certain results on estimation of the density even simple ones; why shouldn't there be similar results for density estimates comparable to those for spectral estimates? It's really a hilarious affair because what is a density estimate but a smoothing of a histogram, right, and it's an example of a more complex situation leading back to a simpler context. In fact, when you go back and take this stuff seriously about the Einstein paper that even goes back 30 years before. So it's an example of how things don't always go the way you think rationally they ought to.

## THE CHICAGO, INDIANA AND BROWN YEARS

**David:** Did you continue to live in Ithaca after you were married?

**Ady:** We were there for a year and went to Chicago. There were no academic jobs and Murray was on his way to accepting a government job; it seemed to be the only thing open at the time. We ended up at our parent's house and then he gets a call there from Chicago. "Would he come?"

**Murray:** Marc Kac, I guess during one of his travels, must have gotten a contact there. I stayed on at Cornell for one year as a postdoc position that was funded by the Office of Naval Research, I think. At that time there was this statistical group at the University of Chicago. It wasn't called a department, but a committee of statistics. Allen Wallis, Jimmy Savage and Charles Stein were members of it. Actually Stein stayed with us for a few months when he first came to Chicago and then he left later on to go to Stanford. That must have been the time after he left Berkeley because of the loyalty oath issue.

**David:** It must have been magic.

**Murray:** After a while, I wanted to leave actually. There were some difficulties. But, that's when I went to Indiana. I guess Julius Blum was at Indiana at that time. In fact, certain aspects of Chicago were very good. Well, there were a number of people that visited. It was a nice aspect of the place, and Grenander visited, and that was the time when we started up doing joint work on the book [see Grenander and Rosemblatt (1957)]. I went to Sweden for two-thirds of a year in 1953. I guess people like Henry Daniels, Mosteller and other such people visited Chicago at that time.

I guess there were difficulties in getting the book with Grenander published. It was initially supposed to be published as part of the University of Chicago series but I guess there's the amusing aspect of how do you get things published and what not and what are the difficulties. They had a very large editorial board which I'm sure had some very good people but in the reviews of the book, some people liked it and some didn't like it. From my point of view, maybe not Grenander's, I thought one had mutually contradictory recommendations. Eventually we had it published by a commercial publisher.

**Richard:** Wiley published the book in the end, right?

**Murray:** Yes.

**Richard:** But this must have resulted in a wider dissemination.

**Murray:** I didn't do it because of that. I did it because I have to admit I did not have any more patience with the Chicago series.

**Richard:** In the end it might have been a better situation all the way around.

**Murray:** Oh, I think looking back, you're right.

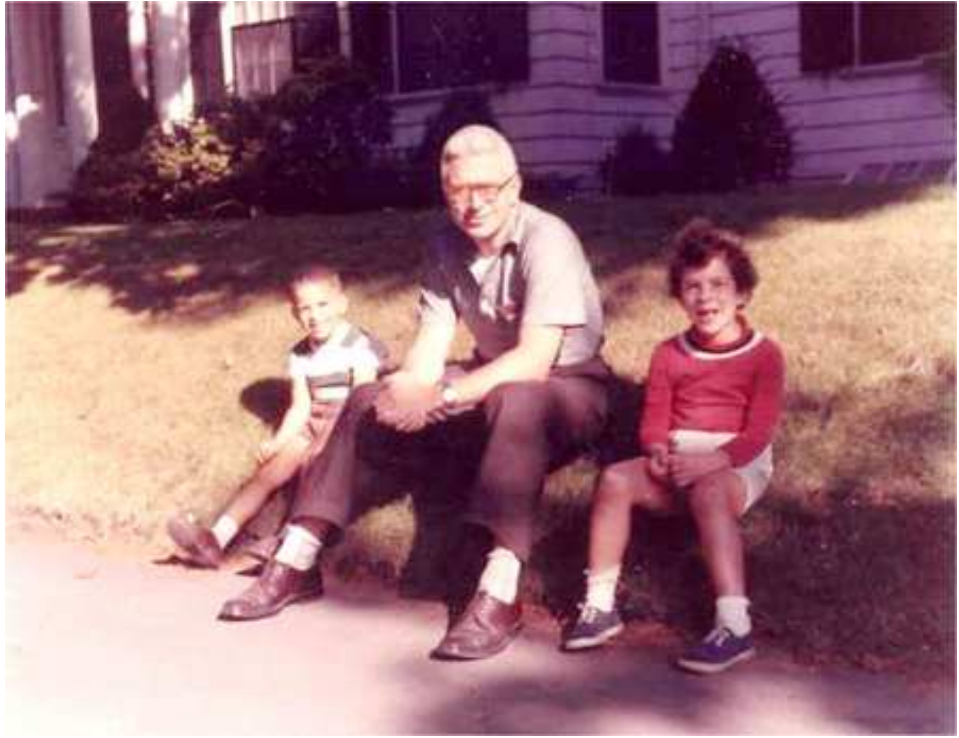

Fig. 4. *In Providence, RI with his two children, Daniel and Karin (1962).*

**David:** You educated me about what it is like concerning refereeing and what it's like when you make a mistake in a paper. I was all upset as I had found a mistake in a paper of mine and you said, "Oh that's your first time?" or some such.

**Murray:** Oh, you were upset about a mistake.

**David:** Yes. I had found a mistake in a paper that had appeared. You calmed me down.

**Murray:** I have had enough mistakes.

**David:** You need someone senior to tell you that when you're young.

**Richard:** It's not the end of the world.

**Murray:** My feeling is that just about everyone running around on earth has found some errors somewhere and hopefully gets it corrected in time or corrected eventually.

While at Chicago, I surely benefited from contact with Bahadur. There's this little paper [23] on density estimation with a little bit on non-parametric aspects at the very beginning of the paper. I certainly benefited from discussions with Bahadur. I remember a dinner where we were invited to by the Bahadurs. I guess we weren't use to spiciness in Indian foods. Initially the shock was my sense of taste. It was overwhelmed. Only gradually did I begin to taste something, but it was very good food.

**Murray:** While at the University of Chicago, I wrote a miniscule paper on economics which actually got published.

**David:** I guess I saw the paper in *Econometrica*, entitled "An inventory problem" [21], and thought it was operations research.

**Ady:** One thing I'll say about Murray: I know a lot of people make a lot of excitement and fuss when they write papers, but when Murray is doing his work he's quiet.

**Richard:** There's no jumping up and down?

**Ady:** No, he doesn't get angry and jump up and down.

**Richard:** He doesn't high-five you when he finishes a paper?

**Ady:** No.

**Murray:** There were some very nice aspects about my times at Chicago, particularly, as I said, with the visitors. One of the visitors I had a nice interchange with was Joe Hodges. We actually wrote two papers. One was a joint paper with Brownlee, which was on the up-and-down method. And the other was rather a cute paper on random walks with Joe, who's really a very bright guy. Did you have any contact with him?

**David:** Oh yes, I did. I remember you saying very early on, when I got to know you, how impressed you were with Joe Hodges. You thought he was a super applied statistician.

**Murray:** Speaking of Joe Hodges, there was a paper, [10], on density estimation unfortunately never published, at least initially.

**Ady:** Didn't you write a paper with the psychologist, Cletus Burke?

**Murray:** We got to meet in Indiana. I worked with him on functions of Markov chains. They seemed to have some interest in economics and psychology, questions on collapsing of states and whether Markovian properties are retained or not.

**David:** I guess it was a hidden Markov process, which has been the rage for many years now. You collapse some states and there's a Markov in the background and you try to learn about it. One can try fitting those things anywhere you can imagine.

**Murray:** Right, I guess the claim is that it's useful. Cletus Burke was a psychologist at Indiana at the time.

**Ady:** Did you do any papers with him?

**Murray:** Yes, there was a paper [7] I wrote with him that appeared in the *Annals of Mathematical Statistics* in 1958.

**Richard:** Murray, what are you looking at over there?

**Murray:** It's just a collection of my papers that I put in bound form so I could remember what I had done. The title of the paper was "A Markovian function of a Markov chain." Oh, I wrote 2 papers with Burke. Another paper was called "Consolidation of probability matrices" [8].

**David:** So he's a psychologist?

**Murray:** He was a psychologist. He was one of these interesting people who was trained as a meteorologist towards the end of WWII. I am very thankful for meeting a bunch of bright characters, including yourselves, along the way.

**David:** You did very well for me when I was a youngster. You too, Richard? Murray, you were a wonderful role model. You remain my academic role model. I hope when I'm 80, I'm talking like this and Lorie is sitting there correcting me.

When did you start doing research in the physical sciences? I first met you at Bell Labs and you were certainly doing it then. In your paper on bandlimited noise and so on, you always seemed to be talking to a lot of engineers.

**Murray:** I don't know when it really started. Partly it may have been already the time at Brown because of the setup at Brown. There was this sort of initial consulting setup together with the professorship there, which was a contact with Bell Labs. That led to a nice contact with David Slepian. Dave and I wrote a paper actually on Markov chains with every n variables independent [33]. I don't know if you ever saw it?

**David:** I know that paper.

**Murray:** It was sort of amusing that there was someone at Bell Labs, Bela Julesz, who was a fellow that was interested in vision and pattern recognition. Partly it focused on the ability of the human eye to see patterns or be able to distinguish between a random assemblage of dots and patterns. That actually led to a joint paper with Dave.

**Richard:** That would be in the early '60s.

**David:** Yes, that's when I met Murray. I was going to ask about the cepstrum analysis because that was what he was working on, the statistics of that. So when you guys came to London, Murray and I had already met. That Brown symposium had a bunch of stuff in it.

**Richard:** Did you attend the Brown symposium?

**David:** No. I had finished at Princeton and I had a postdoc in London. Back then you could take a postdoc and not ever worry about finding a job. It was sort of wonderful because you got a broad perspective, but you didn't expect to make any money, so the two things went together.

**Richard:** Quite a few people went through Bell Labs.

**David:** Bell Labs was wonderful. It was the best job in my life, but now it's been destroyed. It was clearly the best industrial organization and could equal any math or stat department. And you got to work on anything you wanted to work on. There were these tremendous spin-offs; that's what they were counting on. These spin-offs had all these creative people working. Slepian was one of the wonderful guys.

**Murray:** I actually benefited greatly from contact with Dave Slepian. I also met Stuart Lloyd, a very bright guy. While I was there I wrote a paper [29] on narrow band pass filtering.

**David:** That is my favorite paper of yours actually.

**Murray:** I don't know.

**David:** No, I do know! It brought up the engineering in a bright and interesting way. That was a Eureka moment for me.

**Murray:** Now I begin to recall actually, for example, the paper [22] on strong mixing came out while I visited Columbia, even before Bell Labs. I don't know how I came across it but I was motivated to look at some old papers of Serge Bernstein—at least the basic techniques, such as proving the central limit theorem and such, things with the strong mixing by breaking things down to blocks. This goes back to Serge Bernstein or maybe earlier.

**David:** If I could return to cepstrum analysis. When I met you, you were down at the Labs and you were looking at some statistical properties of cepstrum analysis. You know, there was this paper at Brown by Bogert, Healy and Tukey [2], where they did all sorts of stuff. And you were looking at the statistical properties.

**Murray:** That was one of the things that happened at Brown. The Office of Naval Research funded a conference on time series analysis. I and several other people were involved in helping organize it. I edited the proceedings of the conference.

**David:** A climactic moment in time series analysis, that conference. In the book there were important milestones in time series.

**Murray:** It was a nice opportunity to bring things together.

**Richard:** Murray, who attended the workshop at Brown?

**Murray:** Well it was quite an affair. I think I even have a copy of the proceedings. I'll read off the names. There's a paper by Jim Durbin and one by Ted Hannan, one by someone named Lyttkens, (Swedish), M. S. Longuet-Higgins, Gordon Newell, who was at Brown.

**David:** He was a nice man. He was at Berkeley.

**Murray:** Yes and Dave Slepian, and Richard Jones. Then Hasselmann, Munk and MacDonald and Willard Pierson. I don't know if you know that name.

**David:** Yes, the oceanographer.

**Murray:** OK, Monin. I'm not sure if Monin was actually there, but he contributed a paper. Manny Parzen, Enders Robinson. Leo Tick, a fellow called Sirazdinov from the Soviet Union. Then there was this paper, Bogert, Healy and Tukey. This was the cepstrum analysis. And then Walter Freiberger, Roy Goodman.

**David:** He [Goodman] died too young.

**Murray:** Yes, he did. OK, then Jenkins and Kallianpur, Bill Root, Akiva Yalgom, I. J. Good, S. R. Rice, Ted Anderson and myself. It was remarkable to get some of these people.

**David:** It was everybody, Murray.

**Ady:** Didn't somebody say that you mixed too many fields or something?

**Murray:** I don't remember that. That was later. I did some joint work at UCSD with a very bright person in the engineering department, a very good experimentalist, called Charles Van Atta. At that point, ONR and NSF must have contributed a certain amount of money for a conference on "Statistical models and turbulence" [34]. Much of our joint work involved Keh-Shin Lii and Ken Helland, a student of Van Atta. We met up with some reaction from people in the field, versus people outside the field. One person that was supported throughout the years, apparently very well regarded in turbulence, but he was supported individually by ONR and other agencies. I guess he may have felt a competitive aspect.

**David:** I always thought that if one did a study of what would have happened after this conference, you would see an incredible burst of time series activity. Did you make the list of invitees at this Brown workshop?

**Murray:** I'm not sure.

**David:** This was so important, that conference.

**Murray:** I'm sure I contributed but other people had to too. I mean, for example, I wouldn't have known the work of Hasselman, Munk and MacDonald. Later on when I visited England, it must have been partially a Guggenheim fellowship support and also be ONR. I guess that's when we met and we worked on those papers on higher order spectral estimates ([5] and [6]).

**David:** I met you at Bell labs in, like, '61 or '62. It was after that Brown symposium and they had asked you to work on the distribution of the cepstrum estimate. I remember you had some notes because you gave me copies. I don't think you ever wrote a paper on that.

**Murray:** I don't think so; I don't remember it.

I was led to the paper with Dave Slepian. There was the paper on narrow-band noise that comes out of the contact with Bell Labs and partly with Stu Lloyd, then also there were some papers on something on asymptotic behavior of eigenvalues of Toepliz forms [30].

**David:** Slepian was definitely working on that.

**Murray:** There's a variety of people.

**David:** It was a magic place, Bell Labs. It's just so pathetic now.

**Murray:** I think the breakup of AT&T and its support of Bell labs was a catastrophe. They claim that phone calls are cheaper. I think it's a mess today. I don't know what you think.

**David:** I agree totally.

**Murray:** I think that the Bell Labs research group was superb. It was equal if not better than most academic groups. I don't know what influence Tukey had. I never really had any contact with Tukey except at meetings. But I have the feeling he probably had a strong influence on certain aspects of Bell Labs. It was a delightful experience having the contact with Slepian. I used to find Slepian by going down to the Labs. It was a delightful experience having the contact. Slepian was a great person.

**David:** Hamming was another, for example, when you had computing problems. He was another of those very open guys who I had the feeling that if they never wrote a paper in their lives they wouldn't worry about it.

**Murray:** Hamming was sort of an amusing character, sort of a jokester I thought. It was sort of sad leaving Brown. Anyway, I left Brown and was able to get established here. We've been here since then, but I guess most of the group at Brown broke up. The group that I was in was sort of an applied mathematical group, which was focused on some of the classical aspects of applied mathematics, elasticity, plasticity, and fluid mechanics. Well, it's still exists nominally, but it exists with another group that came into place with the group that I was in.

**Ady:** Walter Freiberger comes to mind.

**Murray:** Well, William Prager was the elder statesman of the group initially. Via my daughter one hears an aspect of academic politics. Now it's at a distance and as a retired person, one is relieved of concerns.

## SETTLING INTO UCSD

**David:** You are not retired. Haven't you been doing all this emeritus stuff? I looked you up on the web and that was one of the things I saw; "Mr. President" of the emeriti club.

**Murray:** Well I got involved with that through George Backus. George Backus was president of the emeriti and I guess he didn't have much success in persuading someone else to take over as president so he approached me and persuaded me.

**David:** I got the impression that you did it very seriously. By searching on the web under Murray Rosenblatt, the newsletter *UCSD Emeriti Chronicles* March 2002 comes up. You wrote it.

**Murray:** Oh, that was a nice idea to get reminiscences of people at the beginning of UCSD. A fellow outside of mathematics is persuading people in different areas to write about their background, what happened at the beginning. I think it's a very nice idea and, yes, he persuaded me. I don't know if I did a very good job.

**David:** It's very clear that you were at the start of a very important university.

**Murray:** Oh yes, but it's amusing. I guess some of the administrators at UCSD persuaded someone to come in and write the history. At least my view, you know, it's one of those histories that's greatly distorted. My impression is that what it said about the mathematics department had nothing to do with what actually went on; likewise, some people at Scripps say it seems to be quite a bit of variance from what they remember at Scripps. I've been getting the individual reminiscences of people in different areas. This would be considered a remark on the administrative representation compared to the faculty representations of what took place. Have you ever met up with that?

**David:** Oh, totally! Whenever there's a newspaper article on something I know about, it's totally wrong, so it makes me wonder about the articles I read that I don't know anything about. Can you believe any of those things? But did you have fun when all of this was going on? You were chairman and you were right in the center of a wonderful time.

**Murray:** I was chairman for one year and got out of it; luckily, the person who built up the department was Stephan Warschawski and he brought me as well as other people in. I guess a few years after he came, he had what was called a heart insufficiency so they persuaded me to take on the chairmanship for a year. I guess the notion was to persuade me to continue, but I decided, from my experience from that one year, that one year was enough for me as chairman.

**Ady:** Oh, but you loved it.

**David:** You probably learned how the system worked and so all these wonderful probabilists came, that probably was no accident I'm sure.

**Murray:** No, they actually came while I was away in England. I think people like Ron Getoor, Adriano Garsia and others came and I think that's a tribute to Warschawski, actually. But no, it was great.

I remember composing a five-year plan for the department and I had the feeling that after it was generated it probably ended up in a file cabinet somewhere. The five-year plans, and what later on you saw the department would be faced with, were often mutually inconsistent. So I think five-year plans are always generated and maybe they have an influence and maybe they don't.

**David:** Well, you do something that forces people to think about things, structure things and whether the actual words and details matter, I don't know. I just saw this wonderful university being created in California in San Diego and so on. You're not wanting to take credit for these guys, but I'm sure they were happy to come to San Diego because they had you to talk to and things like that.

**Richard:** What was the sales pitch to bring all these people out to San Diego? It seemed like a risky venture for an established professor to become engaged with starting a new university. I guess you didn't have undergraduates the first year you were here.

**Murray:** Scripps was the basis. It had been here for many years. Initially, the claim was the institution was basically going to be a graduate university. That, unfortunately, disappeared rapidly after a few years. So initially that probably persuaded many people to come. Certain groups may have been encouraged to come together; sometimes that's successful if you get a good group. Sometimes you don't get such a good group and it's not that successful. But often UCSD had success. It was already clear the group was breaking up at Brown. It was clear that certain things weren't that pleasant at Brown.

**David:** It must have been pretty exciting to come out here though: graduate university, southern California, right where the ocean is?

**Ady:** It is right where the ocean is. I think that the chairman Warcharwski was a very good chairman. He made people feel at home.

**Murray:** He was actually the best chairman I ever had.

**Ady:** He and his wife were very sweet. They made us feel very at home.

**Murray:** He was genuinely concerned about building up a good group.

**Ady:** The people from Scripps made us feel very welcome, the ones that had been here.

**Murray:** People like George Backus and Freeman Gilbert used time series analysis to analyze earthquake data. There was always something interesting going on there, and there was this contact with some of the people in engineering. Don Fredkin, a physicist, was a co-author on some papers with John Rice.

**Murray:** We've settled down here. You know, we wandered around a bit. We went to Chicago, and then Indiana and then Brown.

**Ady:** And you were at Columbia for a while.

**Murray:** I was just visiting for a while. It wasn't as long as a year, maybe a semester. I suppose at some time there may have been some sort of negotiation with Columbia. That was already at the time I was thinking of leaving Chicago.

**David:** That was with Robbins and Anderson.

**Murray:** At that time, Robbins and T. W. Anderson.

**Richard:** So the development of probability at San Diego was more by accident. It wasn't a conscious effort?

**Murray:** I think the business of trying to build up the statistics, as well as probability theory, was a conscious effort. People like Richard Olshen and John Rice came. Unfortunately they left, but now we have younger people. Actually, it's curious that there's always this question of contact between the mathematics department and the school of medicine. One of the nice and interesting aspects about UCSD was that initially there was the notion of having professorships that would be tied to academic departments and the medical school. I think Olshen was in one of them. Another person, who was in probability theory, had one of these and was expected to interact with the school of medicine relative to biostatistics.

**David:** People used to talk a lot about the problem of statisticians in a mathematics department. Basic difficulties with salary and things like that.

**Murray:** I think there are difficulties; initially, I was against having a separate statistical group. The statistical group that existed was too small and I had the feeling the likelihood of fracturing would be very great. I think if you have a large enough group, it's a great idea, so statistics still exists to a certain amount. They're about four or five appointments in statistics in the mathematics department, but there are a number of appointments in this community medicine department, maybe 12 people. It's a different sort of affair because, I guess, quite often they are not fully funded by the university. They are funded to a great extent on grants.

**David:** I think they are often supported on contracts not grants, which means they have to agree to do certain things. Creative people don't want to be in that situation.

**Murray:** I think that's a great idea if the problems are interesting and if they can write them up. There are greatest difficulties with their advancement. Often, the doctors give them only marginal credit in papers written jointly. So, if the statistical question is sufficiently interesting, they should write up papers separately. The community medicine department and medical school has had to appoint some statistical types because I suspect Washington demands it now to get some sort of statistical confirmation of medical advances.

**David:** What about the economics crowd? Granger, he's been here for a long time.

**Murray:** I haven't had any great contact with Granger. I've been on committees of their graduate students. I've been to some of their seminars. I feel that economics is a difficult area, really, not in terms of the theories that exist, but in terms of what really happens. Maybe I've always had a certain bias relative to this notion of the "rational man." I think it's so far from what actually happens in real life, I mean, after all the talk of idealized free enterprise. I think it's rare that anyone sees anything like that in actual practice. I should say also I think economics is a very interesting field. However, getting back to what I was saying, one has these indirect investments, you know, in retirement plans and when one looks at what goes on with these companies, with these CEOs, the manipulation of the market, it sounds so different from all these idealized models. I don't know if you have a similar feeling, the both of you.

**Richard:** I don't think about it in quite those terms—I just like to look at the actual time series data!

**Murray:** The actual time series, not locally but globally, has to be influenced sometimes by these affairs. In fact, I think it would be interesting, really interesting, to have someone analyze just what takes place in the stock market due to the intervention of these CEOs. That would be a remarkable affair because I think that would get you a little closer to what sometimes does take place. I'm not sure it would be that difficult to do either.

**Richard:** I found that with some examples—maybe I told you about this—but if you try to fit something like an ARMA model, sometimes you get these noncausal sorts of models that may be suggesting some type of anticipatory action. In examples I've seen, something like the volume shares for Microsoft, you get this behavior. Microsoft is in the news a lot so you can see why this might happen. For other series you may not see it.

**Murray:** I think you are absolutely right. That's not a criticism but an indication that some of the emphasis on causality maybe be overdone occasionally. So my feeling is that economics in certain ways is more difficult than other fields if you really want to get insight into it.

**David:** For example, there is Debreu. He did measure theory, really. He got the Nobel Prize in economics for doing it. There's a whole crowd of them that I don't know about, so they keep away from this real world.

**Murray:** Arrow's work was on this voting system but you know the actual aspects of voting as they really take place are quite different. I'm not decrying any field, it just seems to be a field where occasionally there's a big difference between the idealized models, which are great, but that seem to be taken seriously as compared to what actually takes place.

**Richard:** What do you think about the development of some of these models in financial time series? It seems to be dominating the field right now.

**Murray:** They are helpful as normative models if you want to set some mark, but in some of the well-known cases where you go outside of the assumptions of the model, and the system blows up. For example, the difficulties with the hedge funds, right? It's clear some of the basic assumptions simply weren't satisfied.

**Richard:** Do you have any thoughts about things like GARCH models at all, or stochastic volatility models which are used for modeling financial time series?

**Murray:** I'm really not experienced but I think they are interesting. How successful are the GARCH models?

**Richard:** That's a good question. A lot of people believe that they have severe limitations. On the other hand, they capture some aspects of higher-order moment structures in a certain way because the data are uncorrelated yet dependent.

**Murray:** You've been dealing with some models which aren't nonlinear but capture some aspect?

**Richard:** Yes, the all pass models capture some of the same features but not as cleverly as the GARCH model. People in finance seem to like GARCH models; they seem to work and tap into some essential features, surely not everything.

**Murray:** I'm sure you're right. One of the aspects of economics is that there are large political factors. A firm is going to like it, particularly if you're good at public relations, you can sell your ideas. You can say this is the way to do things. A firm may be willing to pay you a considerable amount of money. And as you say, it's sometimes difficult to see whether what you're selling actually represents what is taking place.

**Richard:** I think there's been some cross-fertilization as well from economics back into mainstream time series. Now we inspect residuals for not just being uncorrelated but for additional nonlinearity such as volatility as manifested by correlations in absolute values and squares of the residuals. We don't only look for these features, but also attempt to model them as well.

**Murray:** But you see there are aspects of that already possibly motivated by another area, long before economics.

**Richard:** This may be something I don't know about.

**Murray:** Because there is the phenomenon of something called intermittency in turbulence. And what is intermittency, but what you refer to from another point of view... what was the term you used?

**Richard:** Volatility?

**Murray:** Volatility. Because if you're looking at time series sometimes it's looking at global structure versus local structure and the intermittency corresponds sometimes to what happens locally. There's this heuristic and theory which people try to formalize that probably originated with Kolmogorov—the notion of what's sometimes called the energy cascade in turbulence, which is looked at from a point of view typically of a spectral analysis. The notion is that the initial energy input is at low frequency and then it cascades down as it's transferred nonlinearly. Of course, turbulence is a three-dimensional phenomenon and economic fluctuations are usually analyzed one dimensionally. And the basic dissipation of energy is supposed to be taking place at the high frequency; this is all sort of heuristic. People have tried to formalize this in various ways but it has its difficulties. Some people look at random solutions of nonlinear equations of motion from a moment point of view. You get an endless sequence of linked moment equations but you try to truncate them at third or fourth order. Isn't this a little reminiscent of the idea behind the GARCH model in a certain sense?

**David:** Yes, one has moved to fourth order moments.

**Murray:** And these ideas have been going around for decades. One wonders whether Engle in his work on GARCH may have been partially motivated by ideas from physics (turbulence).

**David:** Can I try an idea on you, because I think I understand what you're saying? People would bring me time series data and I would think, yes, maybe ARMA and it looks stationary and they are a sort of class of models that I have if someone was asking me and I didn't have the time to do computing and so on, I would say why don't you try the ARMA package? So there would be serious people coming to me and they would show this thing we now call volatility, or intermittency and so on. And now with all this GARCH stuff there's a whole package of programs I can direct people to and they've got residuals and they're forecasting things. So now there's a whole class of wiggly lines developed in your package with Brockwell that we are working with. On a related topic, I feel it must be frustrating to econometricians because they have all these clever ideas with these very difficult problems and often none of them seem to work with their data. So it is the people in other fields who are the ones that take advantage of their clever ideas. It's good they exist, but so be it.

**Murray:** One of the difficulties with economics too is the data. This administration of the younger Bush is not interested in producing good data. They're interested in producing data that agrees with their dogmatic notions.

**Richard:** I'm not sure how hard it is to get the data.

**David:** I get some data.

**Murray:** You can get data, but is the data decent?

**Richard:** I'm not sure that would be the issue for me, it's just the interpretation and the conclusions that are drawn from the data. I don't know if they changed their data collecting methods to adhere to a change in policy.

**David:** They've changed definitions.

**Murray:** Cost-of-living data is obviously, at certain points, highly politically manipulated. Maybe stuff like stock market data is better but we don't know. We have open questions about CEOs and the stock market.

**David:** Clive Granger has come up with a lot of neat ideas. He's been in the statistics we teach and had an impact.

**Murray:** How successful do you think these models like GARCH have been?

**Richard:** I think they do a fair job. The models have problems because there's a dependence issue and there's a heavy tail issue, which are inextricably linked in GARCH models.

**Murray:** The heavy tailed aspect, you mentioned. I was just thinking about these things. I was harping on the effect of CEOs. You have the business of CEOs making their bargain in selling a company. Quite often it's tied up to tremendous benefits to these CEOs once the company is sold, the benefits are a nontrivial fraction of the value of the company. Now it seems to me when things like that happen they've got to have a long effect.

**Richard:** I think it's a hard problem trying to model these types of volatilities directly from the returns, because there's not a lot of dependence in it. We don't have tools to try to find this kind of dependence.

**Murray:** I think you almost have to try to look at some aspect of the system in terms of explanation, as it exists today. I think today we spend a lot of time denying certain aspects of the system claiming it works in certain ways while it doesn't, and we don't quite know why it doesn't work. We know that certain peculiar things take place but we don't know the full extent of the detailed mechanism.

**Richard:** I am not surprised that this would be your approach to this problem. With your background in engineering and the physical sciences, you really want to model the system and understand what's happening there. In other cases, however, one might attempt to model the data without regard to some physical system or the interactions driving the model specification.

**David:** Which are you, Richard?

**Richard:** I think I'm the latter because I'm just not smart enough to figure out the physical system aspect.

**David:** The scientist wants to understand things and so on. Maybe I'm attacking Tukey here, at least his exploratory data analysis. I'm with Murray here. The way you described it, Murray, that's sort of my motto.

**Richard:** I'd like to be like Murray too, but I'm afraid I am not clever enough. I think a lot of these cases in economics—it's just not going to work that way. That's what I think is the beauty of statistics. You can often use a stochastic model for certain phenomena which can do a credible job as a proxy for describing a physical system.

**David:** It would be nice to know who made money doing this from the GARCH stuff let's say, not by writing a book or not by cheating, but they really made money with the straight GARCH stuff.

**Murray:** I'm sure people have made money. I think the GARCH models have become very popular with the stochastic differential equations. I think they set certain levels that are not too bad if you don't take it too seriously. But I think if you push it too far then the thing can blow up.

**Richard:** I think that's the problem with non-statisticians. They take the models too seriously and make more of it than there really is—it's only a model.

**Murray:** Economics is an area where one can make lots of money. I can say even from an academic point of view and I give kudos to the economists for they will get among the highest salaries in the academic area. We can argue how reasonable is it or not but they're effective in selling; in business they're willing to pay a good deal of money.

**David:** What got you doing spectral analysis?

**Murray:** It was basically the contact with Grenander.

**David:** So Grenander was already doing spectral analysis?

**Murray:** I don't know when. He obviously was interested in time series from the beginning and his thesis, I think, was on what sort of discrimination problems there are for time series, particularly parametric forms.... That's the things they credited him for later on, and sieves. Well, the idea of the method of sieves. If you really want to look back, it is really in that paper [12]. You know the background that's something I got to know by contact with Akiva Yaglom. The background on spectrum analysis is really quite amusing historically because an initial heuristic idea was actually in an old paper by Albert Einstein, about 1914, in a Swiss journal [9].

**David:** When Tukey started, there was this radar data and all these different experiences that he had at Bell Labs in terms of the spectrum. Your work had the same feeling about it.

## FAVORITE PAPERS

**Richard:** What is your favorite paper that you've written?

**Murray:** I've never thought about that.

**Richard:** You don't rate them, like a top five?

**Murray:** I'll tell you, the thing that astonishes me is that some of the papers that you think weren't particularly outstanding drew some of the greatest interest and some of the papers that seemed very interesting, didn't. I can't say I'm a good judge.

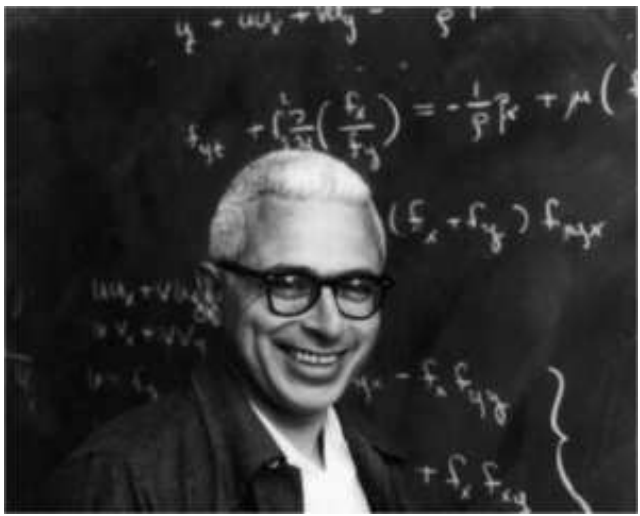

Fig. 5. *Murray at the blackboard (1982).*

**Richard:** This is good to hear from you, actually.

**David:** Some of these papers that people haven't really appreciated, well, 20 years from now all of a sudden they may.

**Murray:** That's an optimistic point of view.

**Richard:** Maybe I can ask you about one with Keh-Shin Lii dealing with higher order spectra and non-Gaussian time series?

**Murray:** Well, I can tell you about one thing that in view of my comments was amusing. It was a paper accepted by the *Annals of Statistics* [16]. For some non-Gaussian models, you cannot estimate the parameters consistently using methods devised under the Gaussian assumption. You had to adapt the procedure. I think we used higher-order spectra estimates to consider some of these parameters. And what was interesting looking back was a positive direction on the part of the *Annals*. The initial reaction seemed to be, from the referee, that this paper is all wrong and should be rejected. Apparently this person had never heard of non-Gaussian processes or the fact that methods devised under the assumption of a Gaussian may not be able to estimate parameters consistently if the process is not Gaussian. You could think intuitively that the Gaussian process works in all cases. We had to write a very detailed reaction. We didn't want to react in a hectic, passionate way; we tried to explain in great detail what the situation was and I guess eventually through an intermediate exchange, one has to say maybe this is a tribute to the *Annals* of that day, we managed to sway the editorial board. You know, there's something to this paper and it's not a situation where all classical techniques work. It's a very common reaction, particularly with some well-established journals which are used to some sort of standardized procedures for certain favorite fields. That if you come up with a procedure which doesn't sound usual or reminiscent of the typical procedures and it doesn't sound familiar, quite often the reaction is there has to be something wrong and it can't be very interesting or whatever. I think for that reason, many journals end up publishing some papers which I'm sure are technically very good but not very interesting and miss out on some of the most interesting papers. I don't know if any of you had any similar experiences.

**Richard:** Often one does receive a number of papers to review from people who are not experts in the area and it's not always easy to make a quick determination about the paper's quality and significance. It can be a difficult chore to filter the papers without much substance from those that are making significant advances.

**Murray:** Right, right.

**Richard:** It's not so easy sometimes.

**Murray:** What happens in many of these cases—forget about any proofs or anything of that sort—does the person read the statement of results and does he try to understand them? In many cases I have found out that is not done, though.

**Richard:** I think you are right.

**David:** I think some people have an attitude that if they are not understanding this by casual reading then the author is not a good expositor.

**Murray:** Probably, or it's a reaction—this sounds so different and I know a reasonable amount about the field that it's got to be wrong since it doesn't sound familiar.

**Ady:** How many years did it take to get that paper forward?

**Murray:** That took about two years.

**Richard:** Is this the one about deconvolution and estimation?

**Murray:** Yes.

**Richard:** That particular paper has had a fairly large impact.

**Murray:** Actually, I think part of it relates to an idea that David had at some time in his paper on the identification of a nonlinear time series system [4]. Part of the engineering community took up these ideas. There was another paper that appeared in the *Annals* later. The associate editor didn't think it was interesting, but a few of the other editors did. I don't think the associate editor understood what was going on.

**Richard:** One thing that seems prevalent in almost all of your research papers is the question, "What if?" You seem to ask this question all the time. What if these conditions aren't true or what happens in this situation? Nonstandard situations often seem to have generated very interesting problems.

**Murray:** Well, I think that is only part of the reason, even though I do play that game occasionally. I don't think that I am well oriented to looking at a question which is super well-defined and wanting to get the best conditions possible for it. Probably even the best conditions depend on how you phrase the affair. It seems to me it's more interesting to look a bit more broadly and see what generally goes on.

**David:** The way I would rephrase what you were saying is that Murray is good at looking for counterexamples. You find something and then you are looking for a counterexample. I don't mean for the result that you prove, but suppose you weaken the assumption, you have your assumption, what is your counterexample?

**Murray:** It's not necessarily looking for a counterexample but there's one case where I eventually looked for a counterexample believing I wouldn't find it, which astonished me and that's a case that's still open. There's some work of Wiener's that he exposited in a little book on nonlinear methods. I think he did some work by himself and some work with Kallianpur [36]. One of the ideas that he had was that under certain conditions, strong enough conditions, you could encode a stationary stochastic process as a one-sided function of a sequence of independent random variables. That's a more stringent version of a problem dealt with by people like Ornstein, Kolmogorov and Sinai. It's sometimes called the isomorphism problem where you get conditions for the process as a two-sided function of a sequence of independent and identically distributed random variables. I guess Wiener had certain conditions that I thought looked reasonable—and maybe something like that is still appropriate—but the way he formulated them the conditions are not sufficient. That led to something like the assumption that the series has a trivial tail field. So I thought I might write something up along that line, but I thought first let me see what happens. I was able to construct a case where those conditions are true but you can't do it as he thought. Today, I think a trivial backward tail field without an additional condition is not sufficient for doing something like that. It is still a very interesting question to find necessary and sufficient conditions for such a representation.

**Richard:** Is this the stationary Markov chain paper?

**Murray:** It is one of the papers around that time, but I am officially so far gone that I don't remember that far back.

**Richard:** Your memory is actually incredible to me. You can even recall going up to the board to do a problem in one of your classes at CCNY!

**Murray:** Some things make a strong impression.

**Richard:** In the 1960 independence and dependence paper [28], you produce an example consisting of squaring a Gaussian process and re-centering by subtracting one.

**Murray:** Yes, that process is not strong mixing since you do not have asymptotic normality of the partial sums.

**Richard:** Not strong mixing since you don't get a central limit theorem.

**Murray:** There was a paper I wrote called "Stationary Markov chains and independent random variables" [27]. And I think that's when I showed that a necessary and sufficient condition for a stationary countable state Markov chain to have such a one-sided representation is that it be mixing. In my paper "Stationary processes as shifts of functions of independent random variables" [25], I generate a Markov chain for which the Wiener construction doesn't yield a one-sided representation, but if you re-encode the process slightly, in an appropriate way, you would be able to. So, as I say, it's still an interesting open question for Markov processes.

**David:** Are you still working on that question?

**Murray:** Every so often I return to it.

**Richard:** What about the paper where you introduce strong mixing and prove a central limit theorem?

**Murray:** The strong mixing was slightly incorrectly formulated. I corrected it in a following joint paper with Blum [1], but the basic argument for the strong mixing for the central limit theorem is given there.

**David:** Richard, there's a paper a fellow wrote about a new mixing condition and Murray then showed the only process that satisfied it was the sequence of independent identically distributed random variables.

**Richard:** Yes, this sounds very familiar.

**Murray:** Richard Bradley then got an extension of my comments. I tried to suggest to this person that there was something strange about his condition. I just don't think he reacted to it and I didn't know what else to do.

**David:** Well that's what Richard did to start us all off in this section of the interview, "What if?"

**Murray:** It was already clear when you have the Russian school with a background of Kolmogorov and these younger people—and remember the Russians had quite a school concerned with turbulence with Monin and Kolmogorov primarily and people like Yaglom who had been interested from that date. So, it was clear already that there were these applications in terms of stochastic methods. They might not have used spectra to a great degree but already people had the exposure to people like Bartlett going back to '46–47. From that point on, he—Bartlett—pursued that area. When I visited London, I visited Bartlett's department, which, actually, I found delightful because Bartlett was a moderately formal guy but a very interesting person to talk to.

**Richard:** I think it's incredible that in 1956 you have these two papers on the central limit theorem under strong mixing, and then this density estimation paper. Did you ever imagine the impact that these two papers would have in the field?

**Murray:** I guess I wrote this paper on the central limit theorem and strong mixing condition while I was visiting Columbia because I'm crediting Columbia at that time, 1956. I've got to say that, actually, the people who really did appreciate it were not people in the States initially, they were the Russians. The paper that obviously took off was that of Kolmogorov and Rozanov [15]. It is the first paper that considers, in the Gaussian case, what are sufficient conditions for strong mixing. Later on it led to this work in Fourier analysis of Helson and Sarason where they got the necessary and sufficient conditions for strong mixing in the case of the Gaussian stationary process. David, these are two colleagues of yours.

I visited the Soviet Union around 1963 and met a number of Russian probabilists. I have pictures of Yaglom, Sinai and Shiryaev.

**Richard:** So they were keen on the central limit theorem under strong mixing paper?

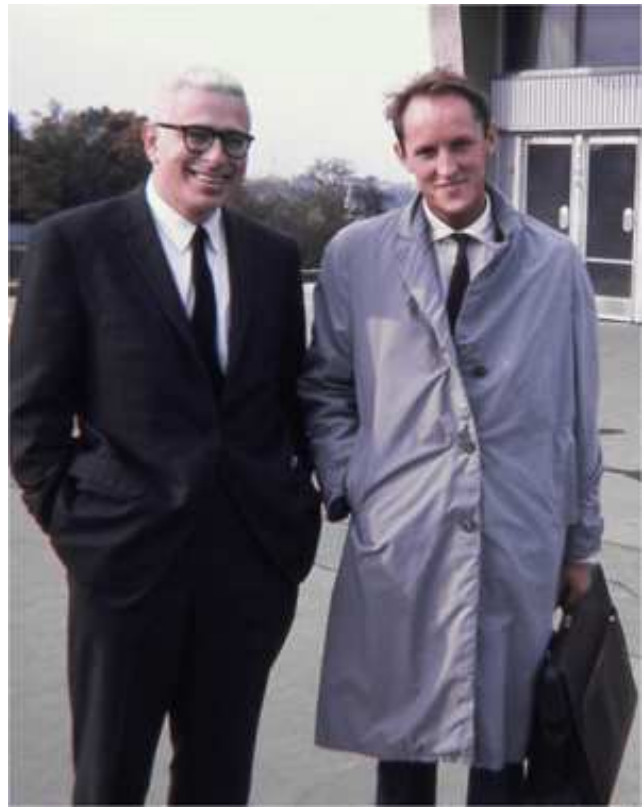

FIG. 6. *Murray with Shiryayev in Moscow (1963).*

**Murray:** Well, the paper, I'm trying to remember when the paper of Kolmogorov and Rozanov appeared. My paper was about 1956 and I think their paper was around 1960. I think there was a whole group of Russians, not necessarily concerned with my version of the mixing conditions, but others like Ibragimov were concerned with mixing conditions suggested by Kolmogorov. You see this in the book [14] by Ibragimov and Rozanov, *Gaussian Random Processes*, a discussion of Kolmogorov conditions and strong mixing conditions. So there was a good deal of activity at that time in the Russian school. There seemed to be more appreciation or more reaction in the Russian school than in the States towards my mixing paper.

**Richard:** Your book [13] with Grenander on time series had a huge impact on the growth and development of time series. It seemed to be way ahead of its time.

**Murray:** Did it? I really don't know how much impact it had.

**David:** I think it was enormous. I had been in these engineers' offices and they had the book—Larry Stark, for example. People like that.

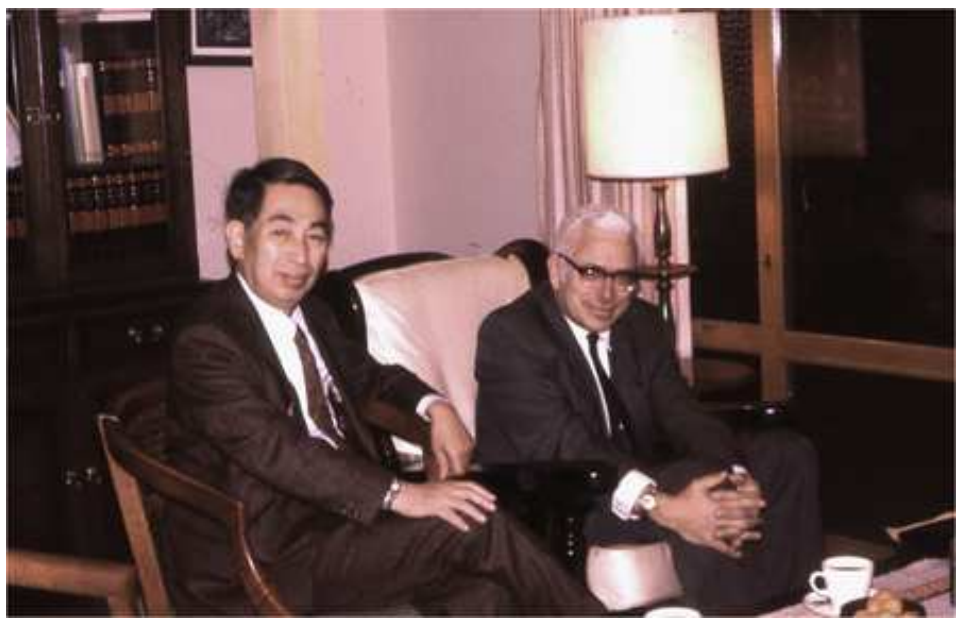

Fig. 7. *Murray with Kiyosi Itô in Kyoto (1976).*

**Murray:** I see. I thought it was unfortunate that it took 3–4 years.

**Richard:** Even so, it just seems like the results in there, including regression with time series errors that Grenander and you developed, are still quite timely.

**Murray:** Oh sure, you have to remember that part of that stuff on regression analysis initially started in some of Grenander's earlier work. I think I carried on some of it to the multivariate case and some papers beyond that. But no, my thought was the book could have quite a bit of an impact. It did take several additional years to get the book in print.

That's why my feeling, relative to papers—and by the way, that is a criticism I would have of the present day of refereeing in what I see in journals. Particularly, in some journals that claim to take themselves very seriously. It seems to me, in present day, the reviewing time, at least in our field, is incredibly long and worse than that when you get the paper back it's not usually refereed really in the old sense—in some cases the person doesn't even read the statement of the results. It seems to me if you receive a paper, you can decide if it's interesting or not rather quickly. If the paper is uninteresting you can send it right back in a fairly short time.

**David:** No one gets harmed when you do that.

**Murray:** That's right, and if you think it may have some interest, then localize the interest and make some positive suggestions or criticism, but get it out in a reasonable time and don't make criticisms on trivia that don't amount to anything.

**David:** Murray, may I agree with that but also disagree? I get more letters "Would you please referee and send a report within 6 weeks" and they are serious.

**Murray:** I get them too, but what I do is I'll say right off the bat I cannot referee within 6 weeks and I could possibly do it in 3–4 months. I think I can do that and I will try to do it, but it seems to me if I can't do it I return the paper to the editor immediately.

**David:** I think a lot of young people get very discouraged at the beginning because this happens. A lot of people never publish their thesis for example. Some people only publish their thesis.

## MISCELLANEOUS MUSINGS

**David:** Murray, I have a question. Richard made some notes on the papers, but maybe you don't want to answer it. I was just curious whether the field of statistics has an impact on the National Academy of Sciences. You were elected a member and that was a very important honor. I am interested in the politics of the American scene and statistics.

**Murray:** Well, I might be able to give some perspective, but there are others who perhaps could provide more insight.

**David:** I was wondering what your experience has been. Was it mainly an honor to be elected?

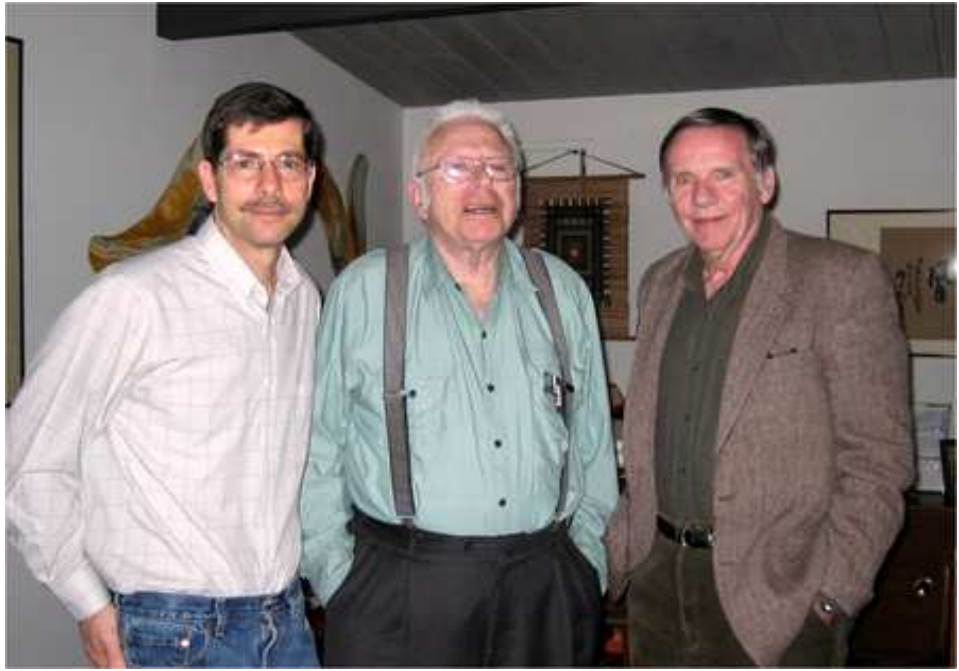

Fig. 8. *Murray with interviewers Richard Davis (left) and David Brillinger (right) (March 2006).*

**Murray:** It certainly is an honor. It gives you the feeling of interaction with different fields. I think Tukey is one of the earliest people elected.

**Richard:** People in probability and statistics, you mean?

**Murray:** Right. Probably the earliest person I know of, obviously a guy who claimed to be a statistician would be Tukey and I would guess that he may have played an important role. You see every so often what the Academy does. There are certain significant fields which are unrepresented or minimally represented so they may increase the number of slots and they may have special nominating committees. I think it's clear there must have been an impetus at certain times for statistics and probability. Doob and Feller were elected in 1957 and 1960; Tukey and Neyman were elected in 1961 and 1963.

I'm a sitting doddering character, but it seems to me that things like time series analysis and the analysis of such data should be taught in statistics departments very broadly. Today they are not, or am I wrong?

**David:** No, no you're right. Manny Parzen has said there are all these very bright statisticians who say time series is hard and seem sort of proud not knowing anything about time series. I have run into this attitude myself.

**Murray:** That's a pity, and I think one wants interaction between time series analysis people, the engineering types and also the types in biology. For example, John Rice and Don Fredkin dealt with certain basic probabilistic models using time series models in helping them analyze the data in a biostatistical context. The context with bioinformatics is great and this leads to increasing types of models. I think it's unfortunate that in many of the departments they restrict themselves to only the classical models.

**David:** I think the bright students are just going to buy the book. They're not going to waste the whole semester sitting in class when something's in a book.

**Ady:** They can get the book and go through it on their own.

**Murray:** I think they lose lots of students who might be very good. Richard, I don't think you decided beforehand what you wanted to go into in terms of ease.

**Richard:** I'm probably not a good example, though, for a typical student.

**Murray:** No, I'm not trying to put you forth as a typical student.

**Richard:** I could have easily done non-commutative ring theory or something like that!

**Murray:** David, what's your own reaction to things like data mining?

**David:** I am going to join them.

**Murray:** Oh.

**David:** The data miners are being sneered at by some of the standard statisticians in some cases.

**Murray:** Are they? I hadn't heard that.

**David:** I think we statisticians want to join the guys doing this stuff in computer science departments or wherever, otherwise our field is going to lose out.

**Murray:** Well there's going to be all sorts of disagreement. Witness the reaction to people saying, you know, you're wasting time if you do have a great deal of computation. Then the notion of exclusiveness arises and you are going to have the same difficulties. It's only if you can have some sort of reasonable interchange that both fields can benefit.

**David:** I think there are a lot of problems out there that people want solutions to. So I don't think that's what going on with the data miners. The statisticians may choose not to join them but I don't think that the data miners are going to resist people coming in and helping them with these big data sets.

**Murray:** It'll be interesting to find that. It's interesting to see what's happened in bioinformatics. I'm sure some part of it may be very bad, but some part of it sounds really quite exciting and interesting. And from a broader point of view maybe more imaginative and involving more interaction with the subject matter. For example, for some diseases they have little idea of what to do. But in cases with powerful statistical techniques they can filter out, from some of these microarrays, a few factors that seem to be relevant. Anyway, what I wanted to say at the beginning is that I actually have had indirectly some association with applications of probability theory and statistics.

**Richard:** I don't think other fields resist statisticians coming in to join them in their research. For the most part, they welcome the assistance—it is often viewed as an add-on to their science. In many cases, certainly not all, it can also be an add-on for statistics as well.

**Murray:** Well, sometimes, it can be an add-on to statistics because I think what happens is, and you see a little of it in the bioinformatics field, at the beginning the initial crude idea is to how you process the data. Some of the initial ideas may have been suggested by statisticians but some may have been dug up by these biochemists themselves. It might be in terms of some typically crude scanning procedure, and asking does the thing look utterly random or can we associate it with something that we recognized before? I think one of the great triumphs of the geneticists has been that they analyze the genetics of these more elementary organisms like fruit flies. Through time, they get to understand a reasonable amount of what certain genes control. One of their main mechanisms has been what one might call a version of pattern recognition. If there's something in the human gene that looks similar to something in the fruit fly genome and if it does, might it control something similar or related to it?

But the trouble is, they shouldn't press it too much. They don't worry about having the exact sequence that you see in the fruit fly genome being reproduced in part of the human genome or something roughly similar, you don't want it too literally, nothing like that. There's been all sorts of transpositions in the genome—God knows what—so something with sort of a rough association.

**David:** Murray, what are you working on now?

**Murray:** A student asked me a question about Markov processes and some conditions I had written on a long time ago [27]. I could answer some of his questions, but not all of them. I tried to answer as well as I could but if there's this lack of clarity in the literature, I should write a short note on the situation. A note [32] appeared recently in *Statistics and Probability Letters.* There are still some further questions on processes with almost periodic covariance functions. These are processes that are not stationary but you can still estimate structure from one sequence using Fourier methods. An open question is what can you do and what not.

**Richard:** What about interactions with some other well-known statisticians or time series people such as Hannan. Did you visit him?

**Murray:** Yes, I did visit him in Australia and I did have interaction with him. We never got to write any papers together but he was a person of great insight. I never found him an easy guy to read, though.

**David:** You know, someone described to me of how Hannan worked. He'd get an idea for a theorem. Then he'd start to try to prove it from the beginning until he got stuck. Then he'd start working on it from the end, working backwards until he got stuck. Then he'd start in the middle and work out in both directions from there. And when all these things connected together, he had his proof.

**Ady:** I understand that.

**Richard:** It was well described—you did a good job.

**David:** Once, I heard that it made things clearer when looking at his papers.

**Murray:** We visited Australia twice. I found it very enjoyable and stimulating.

**Richard:** Did you have much interaction with Whittle?

**Murray:** Whittle is a very talented person. I think he has this personal probabilistic orientation. He's not a person who is noted for rigor but he has developed some extremely powerful ideas via a remarkable intuition. I just find some of it at times incredible.

Hannan was very talented. There are a few of his articles that are clearly written that I could read readily, but most of them I found difficult. Maybe it's due to what you just remarked about.

Whittle, I think had some of these profound insights on how to get decent estimates for time series. They are based on what you might think of as very simple minded, but I think they are very deep. I think they are very powerful because they are so simple.

**Richard:** His idea, the Whittle likelihood, which was developed in the '50s, made this kind of renaissance in the '80s and '90s, probably because of long memory models. For likelihood calculations, exact likelihoods are very difficult to compute, but the Whittle likelihood often gives good results and is much easier to compute, even in complicated models.

**Murray:** Even a good deal of the work of people like Walker and Hannan was really an attempt to try to rigorize some of Whittle's insights, don't you think?

**Richard:** How about moving towards a more controversial subject? The publication of Box and Jenkins' book generated a great deal of interest in ARIMA modeling. They developed a full-service paradigm, often called the Box/Jenkins approach, for carrying out model identification, fitting, prediction, model checking, etc. of ARIMA models. This book seemed to have made quite an impact, especially in business and the social sciences. What are your feelings about this?

**Murray:** I think it did because I think they generated effective means for programs which people could use. It was a strange feeling, I don't know if it had substance, but I had the feeling there was some kind of competitive aspect between Tukey's orientation and the Box–Jenkins' approach.

**David:** I think the two groups were quite competitive from stories I heard. Now, Jenkins was an expert in spectral analysis, and spent time at Princeton, so I guess there wasn't competition with him. For him, Box–Jenkins provided another way to approach the non-stationary time series case.

**Murray:** Sure, sure.

**Richard:** The book was mostly time domain.

**Murray:** It's mostly time domain, that is, time domain versus spectral. But also maybe part of the reaction was in the economics community. There was a group called the Cowles Commission, which existed then at the University of Chicago, and there was a certain amount of real effort by these people to deal with various simple schemes, typically like first-order regressive or first-order moving average. They produced a goodly number of papers on some specific questions there.

However, somehow that work didn't give one the feeling of a general approach. It was only once people starting looking at the general order autoregressive or the general moving average and broader techniques, which I think people like Box and Jenkins did, that there was a flowering effect.

**David:** I think they had a particular audience in mind so they could have been more theoretical, but not for the audience they had in mind.

**Richard:** They also seemed to systematize this whole model system, fitting and that sort of thing, to make it accessible to practitioners. It also seems that if you look at the history of time series, and I'm not a good judge of this by any means, but spectral domain methods dominated for a long period of time and then in the '70s things seemed to shift more toward time domain. Would you agree with that?

**Murray:** I'm not sure about that. I think certainly the time domain theory became much more common. People think more broadly about implementing it. People may get frightened by spectral theory. They claim it's heavy in some sense while autoregressive—moving average models may seem to be simpler. You could immediately write down a set of equations and try to fit the data. Keh-Shin and I published a paper on processes with almost periodic covariance function with mild aspects of nonstationarity where you can still use modified spectral methods. We also wrote an earlier paper that was published in 2002 on processes with spectra on lines, not necessarily parallel to diagonal as is the case for processes with almost periodic covariance function. They also suggest implicitly that if you knew the curve or the locus of the spectra, you could use those spectral methods more generally. In the spectral community, one of the ways of dealing with nonstationarity was the concept of local stationarity. You block off the data in sections and hope you can get a decent spectral estimate in each section and that the spectral estimates change slowly from one section to neighboring sections. And there are people who have tried to formalize the concept. Priestley was one of the people pushing this idea. Dahlhaus also has been pushing an idea like this. Spectral methods are still used often relative to nonstationary geophysical investigations. That used to be one of the ways of dealing with earthquake detection and trying to determine the location of oil deposits using reflectivity properties—all with spectral methods. The geophysicists use it day in and day out and it's turned out to be useful.

**David:** I use the estimated spectrum a lot for residuals. I have a model and now I want to look at something and have it suggest which way the model is inadequate. Other methods do not do so well in my experience.

**Murray:** So you can use spectral methods both for estimation or data processing generally and, well, I presume if you have sufficient data where the things are changing mildly. One of the people dealing with reflectivity problems is Papanicolau. One of the relevant questions is if it's locally stationary, well, what are steps you have to take to get a reasonable estimate. That's a nontrivial question and I think there's a paper in the *Annals* he's written with Mallat that made efforts in this direction and had some interesting comments. Well, in the case of this paper of ours that Keh-Shin Lii and I have written, one of the initial remarks of the subeditor was that anything that isn't stationary, that is nonstationary, you can always use the notion of local stationarity and therefore what's the point of looking at these processes with almost periodic covariance functions. Well, you actually have to use other spectral techniques for these things since they are not locally stationary. You may have spectra on a number of distinct lines and you can't estimate these spectra with locally stationary methods. So, if you had started out with an initial assumption of universal applicability of local stationarity, and if you read that into the paper, you can dismiss the paper. The paper has since appeared in the *Annals of Statistics* [17].

**David:** When you move to want to try to generalize these things to multivariate cases and so on, I think the spectra analysis generalizes quite directly but not the time domain, the Box–Jenkins approach, in particular.

Murray, you've done a fair amount on group theory, and it's scattered through your publications. That's another case where the Fourier approach extends quite directly and is useful, but the analog of the time domain approach hasn't been found yet. Were you doing some things in time series cases having thought of the abstract case and then specializing it to a time series case?

**Murray:** There are the central limit theorems, but also limit theorems in a group- or a semigroup-valued case [30]. Let me see in here. [Looks in a collection of his reprints.] That's one of the hilarious things. The trouble is maybe one's been around too long so you find out when you're looking back you remember the paper but you don't quite remember the details so you have to start looking again.

**David:** Didn't you have a student look at Toeplitz forms as a group?

**Murray:** There was a very bright student that I had at Brown. You may have very bright students but they don't do very much more additional work. And then some people have written work nowhere as interesting as in their thesis, but then they go ahead and do interesting work—more than you might expect.

**Richard:** I was wondering if we can return to a topic that you mentioned a few minutes ago and this is about this attribution issue when people work on stuff in engineering and they don't know the full history.

**Murray:** The question is also what sort of credit do you give to people for their work. What would one say in the density estimation case? Certainly one should refer to the paper of Fix and Hodges. Am I going to say I have exclusivity? That's nonsense. All I can say is some remarks. Some other people claim there is priority by the Japanese group of Akaike, who I'm sure wrote about the same time. In the history of spectral analysis, Einstein had the basic idea at least heuristically. For theorems and conditions, one waited another 30 years or so. So for some results you can say there is a canonical result, but can you say there's a canonical result on density estimation or a canonical result on spectral estimation. I don't think so. Can you?

**David:** No, when you think of density estimation you think of histogram or estimating the bin width for a histogram. They were doing that in the 16th century or so.

**Murray:** Go ahead and smooth it so. Great innovation. Whatever happened, to me it was clear. Look, you can play this game with smoothing, with kernel estimation—you can use any estimation technique. What's sacred about kernel estimates? That's become a big business, hasn't it, with wavelets and this and that, so forth and so on?

What did happen, I have to admit, I was a bit sensitive. You're working in a certain area and there are other people working in that same area and if you're both doing reasonable work, you refer to each other. But it's clear in certain areas that wasn't being done. There were some people or some groups who referred to their in-group and that was it. The rest of the world didn't exist.

**David:** You know how there are some papers people asked to referee and they just wait 2 or 3 years and write their own paper. Lucien LeCam used to talk about this.

**Murray:** That certainly has happened in mathematics a long ways back, you know.

**David:** Laplace, no not Laplace, but Legendre and Newton, I think. Newton was saying that he did it earlier but didn't publish it. Legendre said that meant you can claim to have done anything.

**Murray:** Also there are certain ideas which are in the air for a long time. No one pushes them. There's a time when they really seem interesting. Consider the whole business with the fast Fourier transform. It at least goes back to Gauss in terms of the basic intuitive idea and even the concrete idea. He thought it was a mildly interesting remark. I don't think he ever published anything on it did he? I don't think so.

**David:** In his notebook.

**Murray:** I think it's in his notebooks. And there are people who regenerate various aspects of it. It waited for real interest in computation before people began to talk about it extensively.

**David:** Some people are like this about data. In some fields, the data is everybody's, but in other fields people possess it and release it very slowly. Seismologists, when there's an earthquake involved, they are immediately releasing the data. Also, this biophysicist I work with once said when they are trying to get a molecular structure from different views of a certain particle, the data are not owned by anyone so they are sharing it right away and that's so wonderful.

**Murray:** But that isn't true in other areas. For example, when Keh-Shin and I tried to get some reflectivity data from oil companies, it wasn't available.

**Richard:** Turning to another topic, there are some objects that carry the Rosenblatt name. I'm not sure you were asked about it and maybe you are unaware of the terms.

**Murray:** What's that?

**Richard:** Well there's the Rosenblatt transformation.

**Murray:** The Rosenblatt transformation? I don't know. There's one thing that I found very amusing. I think it was Granger who asked me about one of the first papers I had ever written. It's really a remark on an idea of Paul Lévy's—a transformation on a multivariate distribution that uniformized it.

**David:** Yes, that's it.

**Murray:** I don't know. Have they given my name to that?

**Richard:** I have seen this a lot lately. In fact, I recently came across this term in a PhD dissertation by an applied mathematics student. I looked up the term "Rosenblatt Transformation" on Google and was surprised to see so many hits. Surprisingly, most of the hits are from papers outside of probability and statistics.

**David:** When did Lévy propose it?

**Murray:** I really don't know. If you're interested I can look it up, but can't do it just now.

**David:** I ask because when I was working with fiducial probability, I found a paper by Irvin Segal [35] in which he had this same transformation. He was trying to find things that were pivotal quantities. It reminded me of what Wiener was doing. Wiener was trying to reduce integration and higher dimensional spaces to do integration along the line.

**Murray:** It might be interesting to try to track down all the associations. But my memory is that I think I must have read some section in Paul Lévy's famous book and there was a remark there. I said, oh, why can't you do it two dimensionally or multidimensionally? So that was the gist of my remark and I'm sure it may have appeared earlier, but the basic idea of Paul Lévy goes back at least to the mid '40s, maybe earlier.

**David:** The Cooley–Tukey paper—you know you're talking about the work of Gauss and others. The credit Cooley and Tukey deserve is recognizing the applicability and publicizing it.

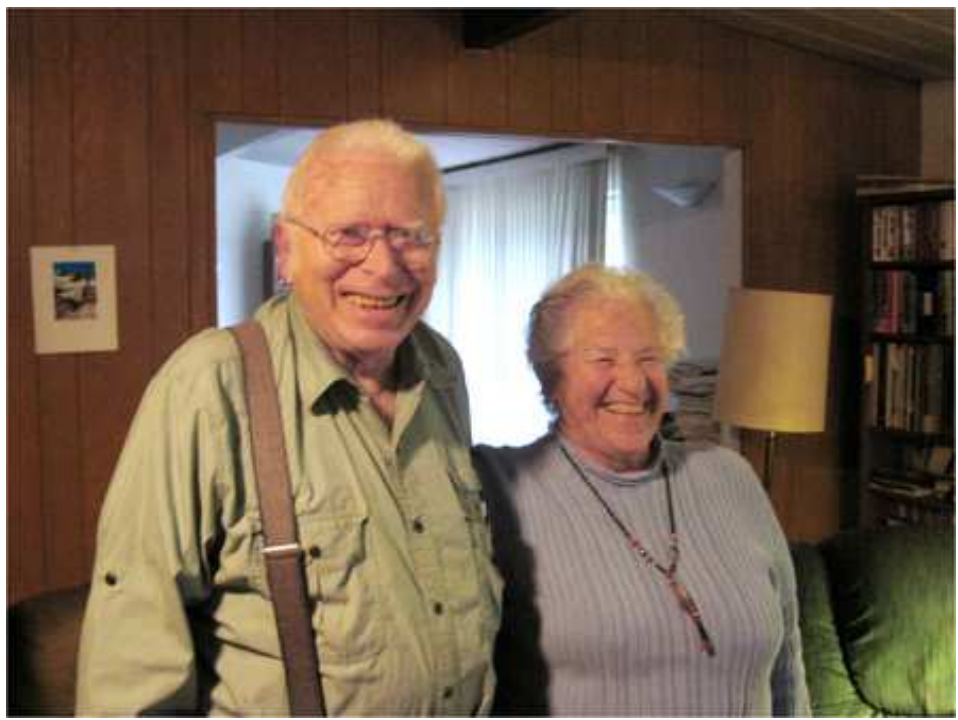

Fig. 9. *Murray and Ady at their home in La Jolla (March 2006).*

**Murray:** Oh, absolutely. I really think, given the interest I have in history, I think scientific history is interesting, but in one way it is off kilter. I think that all people that are associated with some sort of original idea or applicability of it, ought to be given the credit, it would be interesting to trace it through time. But I think that is one thing that is wrong with mathematical crediting. Mathematical crediting looks at such and such a result and often they put a latter-day researcher's name on it and most of the time it is absolutely wrong.

**Richard:** This attribution might have come outside of the mathematics community. It's interesting that Granger had mentioned this to you.

**David:** I'm not surprised that Granger mentioned it because he would think it is an interesting way to grab onto a multivariate problem.

**Murray:** I only found out after the fact that it seemed to be of interest to economists and Granger. He said, well, we found an interesting article of yours. Then when he told me, I thought, gee, what paper could he mean and then I began smiling—oh, it's that paper [20]. Here he was remarking on an idea of Paul Lévy's.

**Richard:** So you hadn't heard this one before?

**Murray:** What, that my name was attached to it? No.

**Richard:** So you can look this up on the web—you'll see lots of references to it. There is another term that you probably already know about, the Rosenblatt process.

**Murray:** That was, well what happened—that's a curious story. I don't know what Taqqu's memory of it would be, but I met Taqqu at some meeting. I guess he must have been a student of Mandelbrot and I don't know if this was his thesis. He greeted me and he essentially said there's something wrong in a paper of mine. But then I thought about it and he explained to me what he thought was wrong with it. I said no, it's not wrong and I think he then looked into it in more detail. You know, that was this business of taking a square of some process, a Gaussian process, with a particular sort of spectrum, which I considered later as an example of long-range dependence. His claim was that some computation was wrong. When I thought about it, it was clear it was not wrong. Luckily I was able to persuade him because when he looked through it himself, then he really got himself involved and obtained some very nice results. He built these results on functions of Gaussian processes. He, on the one hand, and Dobrushin and Major on the other hand, did generate this very nice set of results on limit theorems for such processes. That's why Taqqu called the related process the Rosenblatt process.

**Richard:** I think he actually used it in the title of one of his papers.

**Murray:** I think he may have because I pointed out that it wasn't a mistake. It was worthwhile looking into.

**Richard:** He said that paper really inspired him. I think it was a good example of this paper in the '60s, the independence–dependence paper, where a referee could say this example doesn't really contribute and is not worthwhile to pursue, yet this example inspired him and lots of other people to look at this stuff.

**Murray:** The independence–dependence paper came out as one of the Berkeley Symposium papers. You know, they ask you to write a paper.

**Richard:** So, you never know, this inspired a whole field of people working on this.

**Murray:** Actually, that was one of the more interesting papers I did write. It brought up certain issues and a lot of open questions.

**Richard:** So, are you learning something Ady?

**Ady:** I wrote down the Rosenblatt transformation and the Rosenblatt process.

**David:** Do you ever see some neat result and think, I could have done that?

**Murray:** Well, in fact, I know there are ideas I worked on and got results on and intended writing up and I didn't. From that point of view, I was scooped by someone else because....

**David:** Does that bother you? I guess that's what I was trying ask.

**Murray:** Well, it bothered me a little bit. But it said that I did good work and....

**David:** You seem at peace with life, Murray. Some people it may bother.

**Murray:** No, I think the thing that I may have referred to would bother me. If I had done work in a certain area that should be credited and someone else comes along and says I've done all the work or someone else had done the work and it's simply not true.

**David:** That's not where I was going. I just know you must have countless ideas, used up countless pieces of papers and so on doing things.

**David and Richard:** Ady and Murray, thank you. This has been a treat.

## ACKNOWLEDGMENT

We are extremely grateful to Patricia Osumi-Davis for her patient and splendid work in transcribing the often indecipherable audio tapes of this conversation and related works.